# A JOINT INITIATIVE TO SUPPORT THE SEMANTIC INTEROPERABILITY WITHIN THE GIIDA PROJECT

P. Plini [a], S. Di Franco [a], V De Santis [a], V.F. Uricchio [b], D. De Carlo [b], S. D'Arpa [b], M. De Martino [c], R. Albertoni [c]

[a] CNR, Institute of Atmospheric Pollution Research, Via Salaria km 29,300 Monterotondo st. (RM), Italy – ( plini, difranco, vds)@iia.cnr.it
[b] CNR, Water Research Institute, Viale Francesco De Blasio 5, 70132 Bari, Italy – (vito.uricchio, dario.decarlo, stefania.darpa)@ba.irsa.cnr.it
[c] CNR, Institute of Applied Mathematics and Information Technology, Via de Marini 6, 16149 Genova, Italy – (demartino, albertoni)@ge.imati.cnr.it

**KEY WORDS:** Thesauri, Semantic Interoperability, Information Technology, Web Interfaces, SKOS

**ABSTRACT:**

The GIIDA project aims to develop a digital infrastructure for the spatial information within CNR. It is foreseen to use semantic-oriented technologies to ease information modeling and connecting, according to international standards like the ISO/IEC 11179.
Complex information management systems, like GIIDA, will take benefit from the use of terminological tools like thesauri that make available a reference lexicon for the indexing and retrieval of information.
Within GIIDA the goal is to make available the EARTh thesaurus (Environmental Applications Reference Thesaurus), developed by the CNR-IIA-EKOLab.
A web-based software, developed by the CNR-Water Research Institute (IRSA) was implemented to allow consultation and utilization of thesaurus through the web. This service is a useful tool to ensure interoperability between thesaurus and other systems of the indexing, with, the idea of cooperating to develop a comprehensive system of knowledge organization, that could be defined integrated, open, multi-functional and multilingual.
Currently the system is available in multiple languages mode (Italian - English) and navigation can be done in the following ways: Alphabetical, Hierarchical and for Themes. A full search allows to find any term by searching for the whole term or a part of it and as well as allows to filter the results by themes.
Within a collaborative initiative with the CNR-Institute of Applied Mathematics and Information Technology (IMATI) a SKOS (Simple Knowledge Organization System) version of EARTh was developed. This will ensure the possibility to support the use of the thesaurus within the framework of the Semantic Web in order to be used in decentralized metadata applications.

## 1. THE GIIDA PROJECT

### 1.1 GIIDA in detail

GIIDA (Gestione Integrata e Interoperativa dei Dati Ambientali del CNR) is an interdepartmental project of the Italian National Research Council (CNR). The project is an initiative of the Earth and Environment Department (Dipartimento Terra e Ambiente) of the CNR (Nativi, 2009).
GIIDA mission is "to implement the Spatial Information Infrastructure (SII) of CNR for Environmental and Earth Observation Data". The project aims to design and develop a multidisciplinary cyber-infrastructure for the management, processing and evaluation of Earth and environmental data. This infrastructure will contribute to the Italian presence in international projects and initiatives, such as: INSPIRE, GMES, GEOSS and SEIS. The main goals of GIIDA are:

- Networking: creating a network of CNR Institutes for implementing a common information space and sharing spatial resources.
- Observation: re-engineering the environmental observation system of CNR
- Modeling: re-engineering the environmental modeling system of CNR
- Processing: re-engineering the environmental processing system del CNR
- Mediation: defining methods and instruments for implementing the international interoperability standards.

The project started in July 2008 releasing a specification document dealing with architecture for interoperability and security.
23 proposed pilots from 16 different Institutes belonging to five CNR Departments and from 15 non-CNR Institutions were submitted. These pilot were divided into seven main thematic areas under the responsibility of several CNR Institutes:
• Biodiversity;
• Climate Changes;
• Air Quality;
• Soil and Water Quality;
• Risks;
• Infrastructures for Research and Public Administrations;
• Sea and Marine resources.
Each of these thematic areas is covered by a Working Group which coordinates the activities and the achievements of the respective pilots. Working Groups are called to develop for each area:
1) a specific Web Portal;
2) a thematic catalog service;
3) a thematic thesaurus service;
4) a thematic Wiki;

5) standard access and view services for thematic resources (datasets, models, and processing services);
6) a couple of significant use scenarios to be demonstrated.

**1.2 The Thematic Thesaurus Service**

The Thematic Thesaurus Service is located transversely to the seven GIIDA thematic areas with the purpose of:

- make available through a web interface and a web service, the EARTh thesaurus (Environmental Applications Reference Thesaurus);
- provide support to the different thematic area concerning their present use of terminology for indexing and retrieval;
- interact with the thematic areas in order to develop a system for the collection, analysis, validation of terms candidate for the inclusion into EARTh and for the management of free terms that will be used but that could not be included in the thesaurus;
- interact with the thematic areas for the correct management of specific lists (taxonomies, lists of geological terms, of chemical substances, of geographic terms, etc.) and for their correct use in the metadata creation procedure;
- provide support for a correct use of EARTh terminology during the metadata creation procedure.

The description of databases content through the use of keywords aims to ensure an efficient management of the information. The use of uncontrolled keywords brings to a high lexical ambiguity (polysemy, homonymy) and to the possibility to express the same concept in different ways (synonymy).
Through the development of a semantic network, thesauri ensures that each concept is represented in a unambiguous way.

## 2. THE THESAURUS

**2.1 The semantic model**

EARTh-Environmental Application Reference Thesaurus (Mazzocchi et al., 2007) is based on a multidimensional classificatory and semantic model.
The "vertical structure" of the Thesaurus, built through a deductive (top-down) – inductive (bottom-up) approach, is the fundamental constituent of such a model.
This structure is basically mono-hierarchical. It has been developed according to a tree semantic model and is based on a system of categories. It is organized in a framework composed of different levels and classification knots and comprises hierarchical relationships.
The first level of categories corresponds to Entities, Attributes, Dynamic Aspects and Dimensions.
Entities constitutes "things". Attributes defines character of "things", at least in their static aspects. Dynamic aspects relates to transformations and operations connected to "things". Dimensions identifies the spatio-temporal circumstances where all this is manifested.
Categories are the basis of the semantic model. The vertical structure analyses the meaning of terms. It can be considered as an operative tool that, by providing the interpretation of the primary meaning of the terms and by placing them in the classificatory-hierarchical tree aims to orientate the users towards the most "essential" characteristics of their semantics.
The model envisages the possibility to develop additional arrangements of the terminology. For example, a thematic organization of terms has been elaborated. A theme or a subject is here conceived as a sector of interest that reassembles the terms linked to it (while the tree structure tends to scatter them under their referral logical category). The system of themes as it was conceived, is developed according to the specific needs of the applicative context like the classification of terms for the management of information in the field of environmental policy.

**2.2 The relational structure**

The usefulness of a well-defined and well-structured domain-specific thesaurus for the management of information - also on the Internet - is rather acknowledged.
However there is a widespread opinion that the thesaurus format - as the International Standard conceives it – doesn't completely fit the current needs required by the knowledge organization systems.
One of the main problems posed by traditional thesauri seems to be the fact that they provide a poorly differentiated set of relationships between terms, distinguishing only among hierarchical relationships, associative relationships and equivalence relationships.
It has been also said that since thesaurus relationships are characterized by semantic vagueness, they are not applied consistently. This causes ambiguity in the interpretation and can result in unpredictable semantic structure (Soergel et al., 2004).
The solution that is commonly proposed to overcome these limitations and to enable more powerful searching and intelligent information processing imply the reengineering of "traditional" Knowledge Organization Systems into systems containing domain concepts linked through an extended network of well-defined relationships and a rich set of terms identifying these concepts (Soergel et al., 2004).
In the EARTh project standard relationships will be arranged into richer subtypes, whose semantic content is specified.
This work is particularly useful when dealing with the associative relations (RTs). Typically RTs include a heterogeneous and undifferentiated set of relations, expressing many kinds of association between terms that are not hierarchically based. In EARTh we tried to specify the nature of the relations and to differentiate RTs in subtypes. In this way, the transversal relational structure, which is based on associative relations, will be strengthened. It will be actually developed a knowledge representation model that is net-like structured and could then integrate the taxonomic-hierarchic tree-like model. This will be particularly important in order to deal with the thesaurus subject, the environment, which is a highly complex and interconnected domain. And it will also be useful in order to deal with the networked and barely hierarchical information and knowledge management on the Internet and to better reflect the emerging mental maps of the information searcher. (Trigari, 2003).
The augmentation of thesaurus relationships will ensure a stronger semantic control, also because different relationships can hold each other in check (Fisher, 1998), and open up new application possibilities for information retrieval (Tudhope et al., 2001). Their enrichment and the increased semantic clarification of the relations could enable, for example, a better semantic description of Web resources and guide a user in meaningful information discovery on the Web (Soergel et al., 2004).

**2.3 Terminology content**

EARTh contains at present more than 14.000 terms in English and Italian. Its terminological content is derived from various multilingual and monolingual sources of controlled environmental terminology. First of them is GEMET (GEneral Multilingual Environmental Thesaurus) developed in 1999 by CNR and UBA-Umweltbundesamt, Germany, on behalf of the European Environmental Agency (Felluga, Batschi, 1999). Other sources are represented by internationally acknowledged, more specific domain thesauri.

A further conceptual and terminological enrichment has been defined. Particular attention has been paid in including the terminologies coming from the most recent development of environmental sciences as well as to cover topics that assume strong relevance nowadays, like climate changes, environmental emergencies and disasters..

### 2.4 Multilingualism and cultural diversity

Through managing multilingualism, issues of cultural diversity will also be taken into consideration.

Structural divergences that could concern multilingual thesauri will be considered (Hudon, 1997). Nowadays connection at planetary level is strongly increased. Different cultures and knowledge forms meet on global platforms. We are aware that different cultures hold different visions of the world and this is reflected in the way they organize knowledge too.

We will evaluate in which way the future development of the Thesaurus could integrate a multicultural perspective.

### 2.5 Applications

EARTh represents an environmental semantic map that could be utilized for different purposes. It could be interesting, for example, to evaluate the use of EARTh by considering how it is structured, for dealing with interoperability issues and as a switching tool for mapping among different environment-related thesauri.

Nowadays networked information access to heterogeneous data sources requires interoperability of controlled vocabularies. Thesauri are, in fact, created, with different points of view and can be based on different ways of conceptualization. Their development reflects different scopes and can imply different levels of abstraction and detail. Switching between thesauri, thus being able to create dynamic and semantically based correspondences among different vocabularies is urgently needed (Bandholtz et al., 2009).

## 3. THE WEB INTERFACE

A web-based software is implemented, within a collaborative initiative between CNR IIA and IRSA, to allow consultation and utilisation of thesaurus through the web. This service is a useful tool to ensure interoperability between thesaurus and other systems of indexing, with the idea of cooperating to develop a comprehensive system of knowledge organization that could be defined integrated, open, multi-functional and multilingual.

### 3.1 – Technology

The technological tools applied to the web interface are:
- XHTML 1.0;
- PHP;
- Firebird 1.5;
- AJAX;
- XML;

The PHP (Hypertext Preprocessor) is a general-purpose scripting language that was originally designed for web development in order to produce dynamic web pages. It can be embedded into HTML and generally runs on a web server, which needs to be configured, processing PHP code and creating web page content from it. This language is applied into the system in order to create dynamic pages containing data coming from the query of database (firebird connection/selection). The formatting interface takes place through the using of CSS (**Cascading Style Sheet**) applied in order to improve the way in which the html (or xhtml) documents are visualized and to separate the contents of formatting.

All the pages drawn up are true to the XHTML 1.0. standard.

The AJAX (**Asynchronous JavaScript and XML**) platform, applied to carry out web-interactive applications, is based on a background exchange of data between web browser and server that allows a dynamic updating of a web page without any explicit refilling by the user.

### 3.2 Home Page

The home page of the system, at present accessibile from the following address, http://atlas.dta.cnr.it/GIIDA/, apart from providing information on the team and the project, allows the selection of the thesauri to be consulted (EARTH; EOStem, SnowTerm).

There is also a search function which allows to check the presence of terms within the various thesauri.

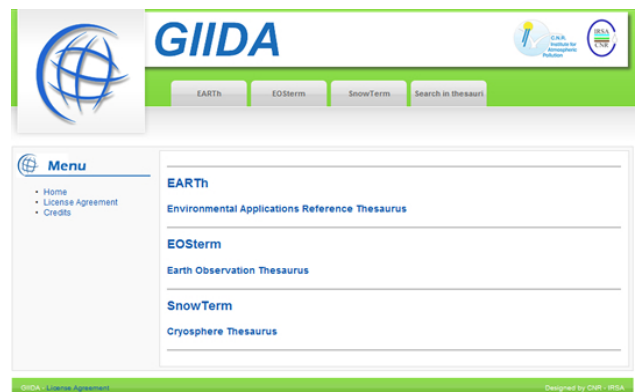

Figure 1: Interface of GIDA – home page

### 3.3 Features of the web interface

Currently the system is available in multiple languages mode (Italian - English) and navigation can be done in the following ways:
- Alphabetical
- Hierarchical
- Thematic

The system foresees a complete navigability into all the terms, each other connected by any semantic relationship. The page of the search is represented as in the following figure.

Figure 2: Search Form

The ways of searching are:
- <u>search phrase exact</u>, in which you seek the complete term in the thesaurus;
- <u>search phrase any</u>, in which you seek the terms containing the wanted word;
- <u>begins</u>, in which you seek all the terms beginning with the wanted word;
- <u>terminates for</u>, in which you seek all the terms ending with the wanted word.

Each search can be filtered by indicating from one to three themes so that only the terms included in the suitable themes will be shown.

For each term included in the database the system will provide the following information:
- the term in the English and Italian translation and the corresponding definition, if it is present;
- the synonyms of the term;
- the terms linked to the visualized term with the indication of the kind of relationship;
- the broader terms;
- the narrower terms;
- the themes linked to the term;
- the link to thesaurus GEMET in order to visualize the term immediately in the EIONET GEMET thesaurus.

Figure 3: Term information displayed on the Web

## 4. THESAURUS AND SKOS

An advanced version of EARTh has been developed within a collaborative initiative with IMATI-CNR in order to further encourage its adoption and exploitation from third party. In particular the aim is to make the thesaurus accessible for the Semantic Web community and for all word wide users who are interested in searching environmental resources by Web of Data.

The activity has required an investigation and design of the thesaurus schema and technology to be adopted. Solution has been identified according to semantic web\W3C recommendations: the thesaurus has been encoded in the standard model SKOS (Simple Knowledge Organization System) (SKOS, 2009) and linked data best practices (Berners-Lee 2006) (Bizer, Cyganiak, Heath, 2007) has been selected to publish the thesaurus in machine understandable format exploiting existing protocols in the web.

We assume that the requirements for a good exploitation of a thesaurus from third parties are:
- to browse its content as HTML Page by a web client;
- to get a semantic web compliant representation of the terms;
- to refer to the thesaurus content without to replicate it, so third parties can extend the thesaurus adding concepts, lexical representations, and so on, referring to terms data already published;
- to interlink the thesauri with other thesauri already published as linked data.

The combination of SKOS as vocabulary to express the thesaurus content with RDF (RDF, 2004) as encoding and the publication of thesaurus according to the linked data best practices, which associates to each concept ID a http dereferentiable URI published on a server allows to meet all the previous requirements.

### 4.1 Methodology

The methodology adopted is characterised by two main activities:
1. Resource translation in SKOS: EARTh content has been encoded in SKOS;
2. Publication: EARTh has been made available on the web according to the Linked Data Best Practices.

In particular EARTh has been published in SKOS taking advantage of D2R server, which is provided by the free university of Berlin (D2R, 2009). D2R Server is a tool for publishing relational databases on the Semantic Web. It enables RDF and HTML browsers to navigate the content of the database, and allows applications to query the database using the SPARQL query language.

The implementation process, illustrated in Figure 4, is characterised by the following actions:

A. *Import of Earth DB in MySQL* Server. D2R work in principle with any Relational Database, however some managing facilities are provided for MySQL. For this reason, the first action has been to import EARTh database in MySQL Server;

B. *Creation and extraction of a view of the EARTh content.* Mainly because of performance reasons and to simplify the mapping with the EARTh data model, a view of the EARTh content has been built implementing a proper store procedure;

C. *Mapping between the extracted view of EARTh and SKOS.* A mapping between the view of EARTh and SKOS has been performed, defining a configuration file according to the (D2R Mapping Language D2RQ);

D. *Set up of the server D2R.* The server D2R has been set up at a given URL.

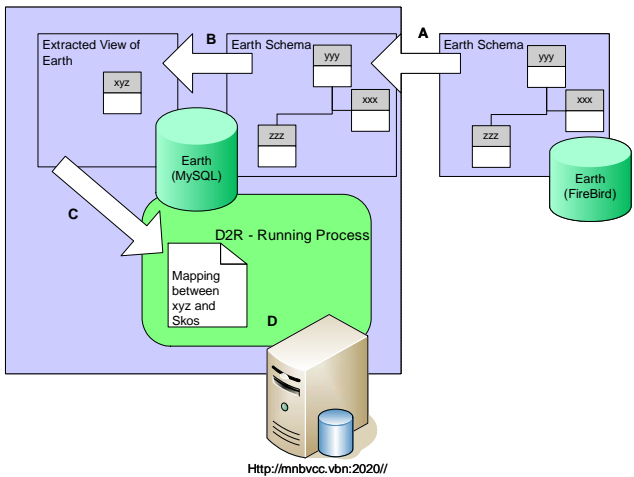

Figure 4: Methodology to encoding and publish EARTh in SKOS.

### 4.2 Result

All the EARTh concepts encoded in SKOS are available on a IMATI server at the URI http://linkeddata.ge.imati.cnr.it:2020/directory/EARTh. Figure 5 and Figure 3 illustrate how it looks like when one of its concepts is accessed by an http browser and RDF browser respectively.

Currently the published version of EARTh includes: preferred labels in Italian and English, brother and narrows terms, related terms, definitions (it, en), mapping to GEMET concepts.

Earth publication in SKOS/RDF with D2R server according to Linked data best practices allows:

- to browse EARTh as HTML Page by a web client (e.g. Internet Explorer) as well as to get a Semantic Web compliant representation of Earth in order to resolve the URI by "semantic" web clients (e.g., Tabulator (Tabulator, 2009)) and to query the thesaurus content by third party applications by SPARQL;
- to improve the effectiveness of an extension of EARTh: it is possible to add and link new concepts, lexical representations etc. to EARTh without to replicate it;
- to create a connection among other thesauri according to the linked data best practices. For example it is possible to make an interlinking of EARTh with other thesauri published by third parties as linked data (e.g., GEMET).

Figure 5: Screenshot of EARTh term "inadequate rescue" on D2R server dereferencing by an HTTP browser (e.g., Firefox, Internet Explorer).

Figure 6: Screenshot of EARTh term "inadequate" rescue on D2R server dereferencing by an RDF browser as Tabulator.

### 4.3 Example of EARTh exploitation for Nature Conservation

EARTh encoded in SKOS has been exploited for the realisation of a Common Thesaurus for Nature Conservation within the European project NatureSDIplus. A project result has been a Common thesaurus Framework for Nature Conservation where to share some general content thesauri (e.g. GEMET, EARTh) and domain knowledge organization systems (e.g. EUNIS Species, Habitat...) for specific geographic data themes according to INSPIRE directive such as Protect Site, Biogeographical Region, Habitat & Biotopes, Species Distribution. A thesaurus may be just added or if needed linked with other thesauri inside the framework or with other thesauri which are not contained in the framework. The result is a dynamic environment where a new thesaurus can be added or further extended and two different thesauri may be interlinked.

A general view of the Common Thesaurus Framework is illustrated in Figure 7. It is contains EARTh as general purpose thesaurus and other specific knowledge organization systems for the four data themes. In particular EARTh has been interlinked both with other internal KOS (e.g. DMEER of biogeographical regions) and with external thesaurus (e.g. a linkage to GEMET is included even if GEMET is not directed published as part of the framework).

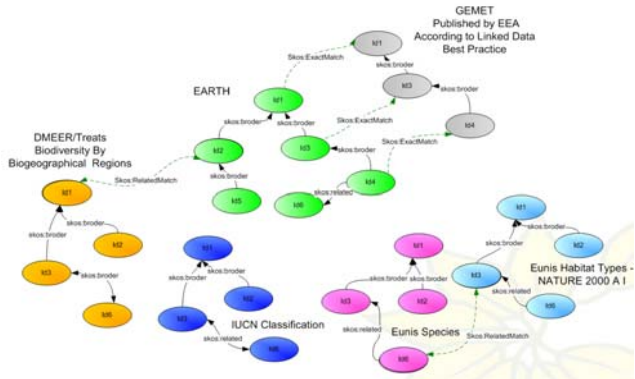

Figure 7: Common Thesaurus Framework for Nature Conservation: EARTh exploitation.

## 5. REFERENCES

**References from Journals**:
Fisher, D.H., 1998. From Thesauri towards Ontologies?. *Advances in Knowledge Organization*, Vol. 6, pp. 18-30.

Hudon, M., 1997. Multilingual thesaurus construction: Integrating the view of different cultures in one gateway to knowledge and concepts. *Knowledge Organization*, Vol. 24 (2), pp. 84-91.

Mazzocchi F., De Santis B., Tiberi M., Plini, P., 2007. Relational Semantics in thesauri: Some Remarks at Theoretical and Practical Levels. *Knowledge Organization*, vol. 34 (4). pp. 197-214.

**References from Books**:
Smith, J., 1989. *Space Data from Earth Sciences*. Elsevier, Amsterdam, pp. 321-332.

**References from Other Literature**:
Felluga, B., Batschi W.D. (Eds.), 1999. GEMET, General European Multilingual Environmental Thesaurus. Version 2.0. European Environmental Agency, Copenhagen.

Bandholtz T., Fock J., Legat R., Nagy M., Schleidt K., Plini P., 2009. Shared Terminology for the Shared Environmental Information System. Environmental Informatics and Industrial Environmental Protection: Concepts, Methods and Tools, 23rd International Conference on Informatics for Environmental Protection., Volume 1. Shaker, Aachen, pp. 123-127.

**References from websites**:
Nativi S., 2009. The GIIDA (Management of the CNR Environmental Data for Interoperability) project. Geophysical Research Abstract, Volume 11, EGU2009-3425. http://meetingorganizer.copernicus.org/EGU2009/EGU2009-3425.pdf (accessed 19 nov 2009)

Soergel, D; Lauser, B.; Liang, A.; Fisseha, F.; Keizer J.; Katz, S., 2004. Reengineering Thesauri for New Applications: the AGROVOC Example. "Journal of Digital Information", Volume 4, Issue 4. Article No. 257 (2004-03-17). http://jodi.ecs.soton.ac.uk/Articles/v04/i04/Soergel/ (accessed 20 Jan. 2005)

Trigari, M., 2003. Old problems in a new environment. The impact of the Internet on multilingual thesauri as research interfaces. In: MULTITES CONFERENCE (London: September 29-30, 2003). http://www.multites.com/conference03.htm. (accessed: 8 Nov. 2004)

SKOS, 2009 SKOS Simple Knowledge Organization System, Reference, http://www.w3.org/TR/skos-reference/, (accessed: 10 Nov 2009)

Berners-Lee, 2006. Berners-Lee, T. (2006) Linked Data http://www.w3.org/DesignIssues/LinkedData.html, (accessed: 1 Nov 2009)

Bizer, Cyganiak, Heath, 2007. How to Publish Linked Data on the Web, http://sites.wiwiss.fu-berlin.de/suhl/bizer/pub/LinkedDataTutorial/20070727/, (accessed: 10 Nov 2009)

D2R Server, 2009. D2R Server, Publishing Relational Databases on the Semantic Web, http://www4.wiwiss.fu-berlin.de/bizer/d2r-server/, (accessed: 10 Nov 2009)

D2RQ, 2009. D2RQ mapping language, http://www4.wiwiss.fu-berlin.de/bizer/d2rq/spec/20090810/, (accessed: 10 Nov 2009)

Tabulator, 2009 http://www.w3.org/2005/ajar/tab, (accessed: 10 Nov 2009)

RDF 2004, RDF/XML Syntax Specification (Revised), Dave Beckett, Editor, W3C Recommendation, 10 February 2004. Latest version available at http://www.w3.org/TR/rdf-syntax-grammar/, (accessed: 10 Nov 2009)

Tudhope, D.; Alani, H.; Jones, C., 2001. Augmenting thesauri relationships: Possibilities for Retrieval. "Journal of Digital Information", Volume 1, Issue 8. Article No. 41, (2001-02.05). http://jodi.ecs.soton.ac.uk/Articles/v01/i08/Tudhope/ (accessed: 20 Jan. 2005)

### 5.1 Acknowledgements

The activity performed by IMATI has been partially supported by the European Commission within the NatureSDIplus project of eContent Programme.